%% file: dasu.tex
\documentclass[12pt]{article}
\usepackage{pic03}
\usepackage{hyperref}
\usepackage{url}
\usepackage{graphicx}

\def\BaBar{{\sc BaBar}}

\input babarsym
\input symbols

\begin{document}

\title{\bf PROBING THE STANDARD MODEL WITH ELECTROWEAK PENGUIN 
             $B$ DECAYS}
\author{
Sridhara Dasu        \\
{\em Department of Physics, University of Wisconsin, Madison, WI 53706}}
\maketitle

%
% photograph of author
%  This is where we will insert a photograph. To see what it would look like,
%  uncomment the following lines.
%
%\begin{figure}[h]
%\begin{center}
%
% include photograph for proceeding version
%
%\includegraphics
%[height=4.5cm]{einstein.eps}
%
% insert a fixed vertical spacing instead for the ArXiv preprint
%
\vspace{4.5cm}
%
%\end{center}
%\end{figure}

\baselineskip=14.5pt
\begin{abstract}
Recent branching fraction and asymmetry results of Electroweak Penguin
$B$ decays from \BaBar, Belle and CLEO experiments are reviewed. While
these branching fractions are consistent with the Standard Model
expectations and are being used to extract heavy quark model parameters
and CKM matrix elements, the asymmetry results are just becoming
sensitive to observe any new physics effects.
\end{abstract}
\newpage

\baselineskip=17pt

\section{Introduction}

It has long been observed that flavor changing neutral current (FCNC)
transitions, e.g., $b \rightarrow s \gamma$, $b \rightarrow d \gamma$
and $b \rightarrow s l^+ l^-$, are suppressed in nature. In particular
the exclusive electroweak decays, $B \rightarrow K^* \gamma$ and $B
\rightarrow K^{(*)} l^+ l^-$ are known to have branching fractions at
or below $10^{-5}$. The Standard Model, by its very construction, does
not allow tree level FCNC. The rarity of these decays is naturally
explained due to the requirement of higher order loop (penguin and box)
diagrams . New physics models must also contend with stringent
restrictions on FCNC. However, new particles, e.g., charged higgs
bosons of models with two higgs doublets, and charged scalars of the
super symmetric (SUSY) models do enter these loop diagrams modifying
the $b \rightarrow s X$ and $b \rightarrow d X$ amplitudes.
Interference between the Standard model amplitude and these new
amplitudes can manifest as an increase in the branching fractions or
more subtly in increased direct CP violating or isospin violating
asymmetries.

The CLEO experiment, which collected 9.1 $fb^{-1}$ on $\Upsilon$(4S)
resonance and 4.4 fb$^{-1}$ 60 MeV below it, was first to measure many
of these decays. The new B-Factory experiments \BaBar\ and Belle are
now producing results using new techniques with much larger
luminosities, up to 113 (140) fb$^{-1}$ on resonance and 12 (18)
fb$^{-1}$ off resonance for \BaBar\ (Belle). In this paper we review
the status and examine the prospects of electroweak penguin $B$ decay
measurements. We also discuss the extraction of heavy quark model
parameters and CKM matrix elements from these measurements.

\section{Experimental Techniques}

The radiative electroweak penguin $B$ decay signals are difficult to extract
due to large continuum and combinatorial background in $B\bar{B}$ events. 
The background is composed of initial state radiation photons, and photons 
from neutral meson ($\pi^0, \eta, ...$) that have been misidentified as single 
photons. The leptonic electroweak penguin decays are mainly combinatorial
arising from double semileptonic decays of heavy mesons. Although the
backgrounds in this case are lower, the signals are also expected
to be suppressed compared to the radiative decays making these 
measurements challenging.

The Standard Model expectations for the quark level decays are
calculated to high accuracy. However, it is necessary to use
approximate theoretical models that take into account the
non-perturbative hadronic effects to calculate B meson decay processes
that we measure in the laboratory. These calculations are most reliable
when inclusive measurements are made. Unfortunately, making inclusive
measurements is experimentally more difficult. In order to suppress the
backgrounds three different techniques have been used thus far. The
CLEO experiment collected substantial continuum data to make
statistical subtraction of non-B background. All experiments are
measuring radiative B-meson decays in as many exclusive hadronic
states as possible. The higher statistics \BaBar\  and Belle experiments are able
to use a partially reconstructed second B meson in the event, e.g., a
semi-leptonic B decay, to suppress the continuum background.

\section{$b\rightarrow s \gamma$ measurements}

The SM $b\to s\gamma$ branching fraction is predicted to be ${\cal
B}(b \rightarrow s \gamma) = (3.73 \pm 0.3) \times
10^{-4}$~\cite{misiak} at the next--to--leading order (NLO). The
present theoretical uncertainty of $\sim 10\%$ is dominated by the mass
ratio of the $c$--quark and $b$--quark and the choice of the
renormalization scale. New Physics contributions with e.g. charged
Higgs exchanges or chargino--squark loops are expected to be at the
same level as the SM ones. Unfortunately, initial measurements have already
ruled out possibility of discovering any dramatic new physics effects here.
However, \CP asymmetries do provide a stringent test of the SM. While 
small in the SM ($\leq 1\%$) \cite{soares} the \CP asymmetries can reach 
$10$--$50\%$ in models beyond the SM~\cite{kagan}.

The photon energy spectrum in $b\to s\gamma$ is used to understand
the hadronic effects as it only depends on the parameters defining the
structure of the B mesons. For instance, the moments of the photon
energy spectrum are used to measure the Heavy Quark Effective Theory
(HQET) parameters which determine the $b$--quark pole mass
($\overline{\Lambda}$) and the kinetic energy
($\lambda_1$)~\cite{ligeti}. These parameters are needed to obtain a
precision value of $| V_{cb} |$ from the $b \to c \ell \nu$ inclusive
rate, and $V_{ub}$ from $\B \to X_u \ell \nu$. 

At the lowest order in $\Lambda_{QCD}/M_B$, the $B \rightarrow X_s \gamma$
photon energy spectrum, where $X_s$ refers to inclusive strange
hadronic states, is given by a convolution of the parton level $b \to s
\gamma$ photon energy spectrum with the light--cone shape function of
the \B\ meson, which describes {\cal all} $b$ to light--quark
transitions. At the same order in $\Lambda_{QCD}/M_B$, the $\B \to X_u
\ell \nu$ lepton energy spectrum is given by a convolution of the
parton level $b \to u \ell \nu$ lepton energy spectrum with the same
shape function~\cite{vub}. Corrections up to the next order of
$\Lambda_{QCD}/M_B$ are currently the subject of active
investigation~\cite{bauer}.
%
%%%%%%%%%%%%%%%%%%%%%%%%%%%%%%%%%%%%%%%%%%%%%
\subsection{Inclusive $b\to s \gamma$}

Two experimental approaches have been used to measure the inclusive
rate for the $\b\to s \gamma$ process. 

The ``fully inclusive'' method measures the high energy photon
spectrum without identifying the hadronic system $X_s$. Continuum
backgrounds are suppressed with event shape information, and then
subtracted using off--resonance data. \B\ decay backgrounds are
subtracted using a generic Monte Carlo prediction, which is
cross--checked with a $b\to s\pi^0$ analysis. \BaBar~\cite{BaBarbsg}
has presented a preliminary result from a fully inclusive analysis in
which the ``other'' \B\ is leptonically tagged to almost completely
suppress the continuum background reducing the need for large
off-resonance data, and exploiting its high statistics on-resonance
data. These methods do need to tackle the \B\  decay background.

A ``semi--inclusive'' method, which measures a sum of exclusive $B\to
X_s\gamma$ decays, has been used by both \BaBar~\cite{BaBarbxs} and
Belle~\cite{BELLEbxs}. The hadronic $X_s$ system is reconstructed by
\BaBar~(Belle) in 12 (16) final states with a mass range up to 2.40
(2.05)~\gev. This includes about 50~\%\ of all $b\to s\gamma$ final
states. Continuum and \B\ decay backgrounds are subtracted by a fit to
the beam--constrained \B\ mass in the same way as in an exclusive
analysis. The result is extrapolated to obtain inclusive branching
fraction using Monte Carlo simulations.

\begin{figure}[hbtp]
 \begin{minipage}[c]{7cm}
\hbox to\hsize{\hss
\includegraphics[height=5cm]{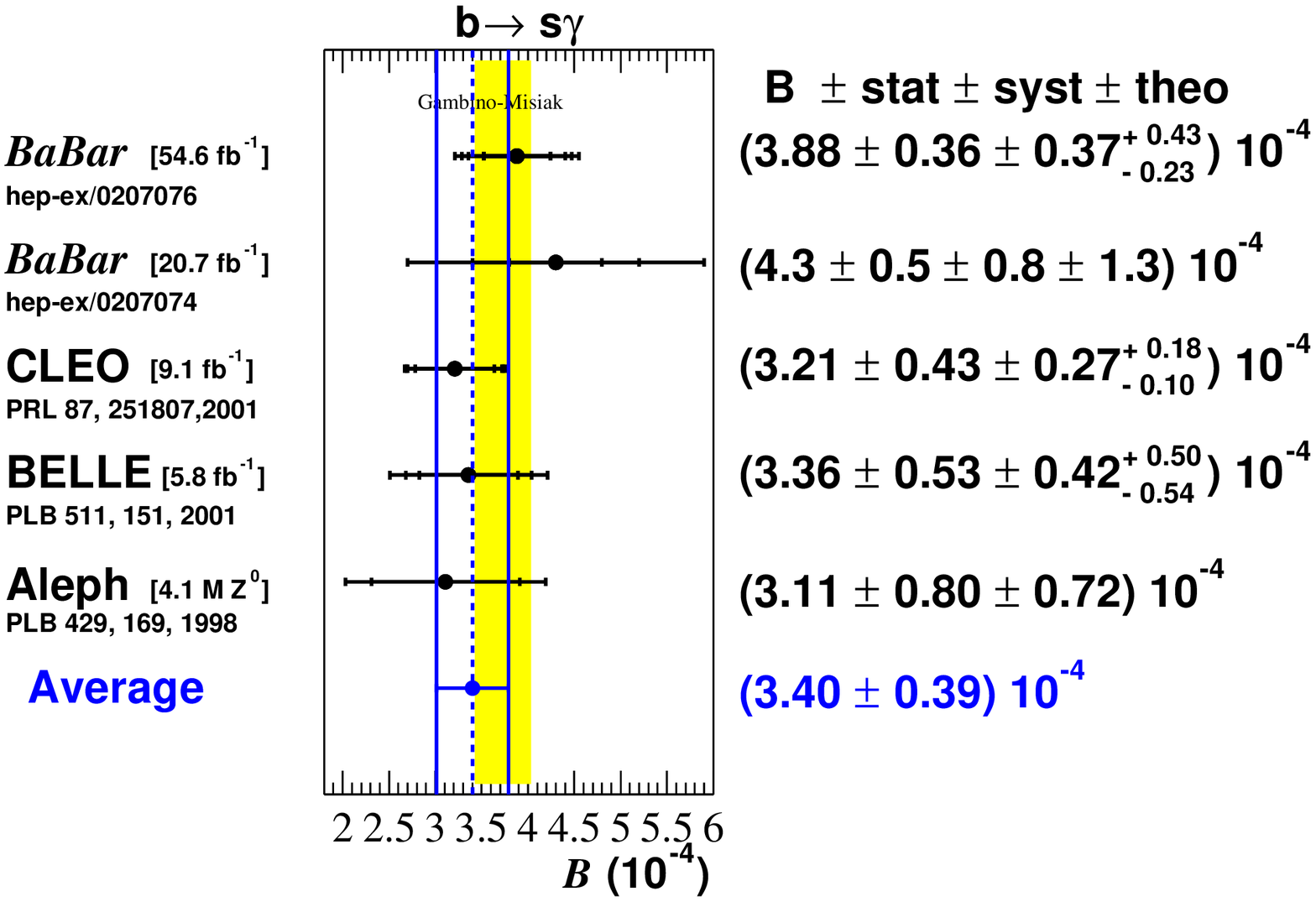}
\hss}
\caption{Summary of $b\to s\gamma$ branching fractions. The shaded band 
shows the theoretical prediction described in Ref.~\cite{misiak}.}
\label{fig:bsgBF}
\end{minipage}
\hspace{1cm}
 \begin{minipage}[c]{7cm}
\begin{tabular}{c}
\includegraphics[height=3 cm]{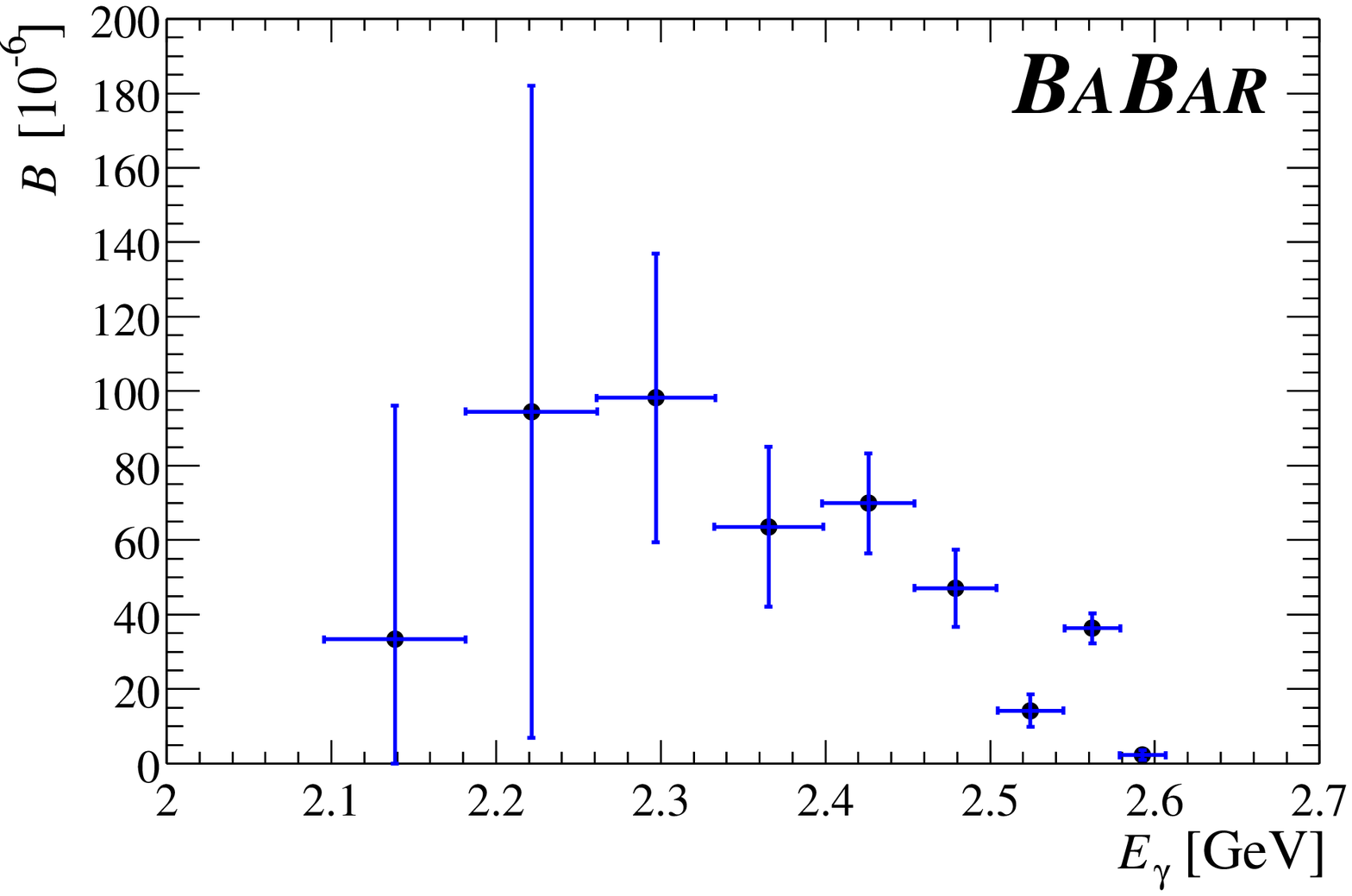}\\
\includegraphics[height=3 cm,width=4 cm]{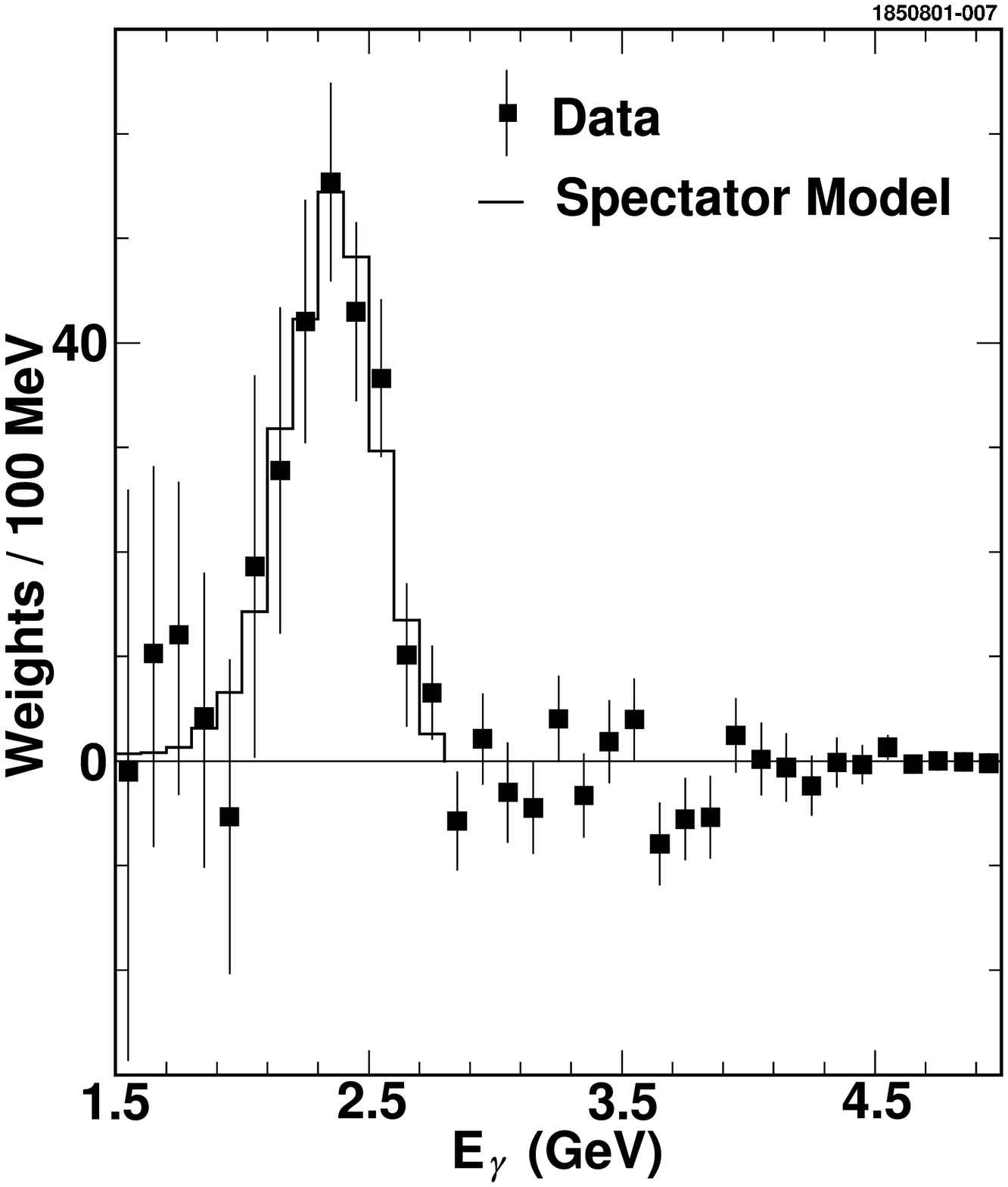}\\
\end{tabular}
\caption{\BaBar~\cite{BaBarbxs} (upper plot) and CLEO~\cite{CLEObsg} (lower plot) photon energy spectrum in $b\to s\gamma$
decays.}
\label{fig:inclusive}
\end{minipage}
\end{figure}

CLEO~\cite{CLEObsg} has published a measurement combining several
methods each with a different technique to reduce the background. This
measurement is still the best single result. Figure~\ref{fig:bsgBF}
summarizes the measurements of the $b\to s\gamma$ branching fraction.
The theoretical error from the extrapolation of the inclusive rate from
the measured energy range to the full photon spectrum is quoted. CLEO
has a lower threshold (2.0~\gev) than \BaBar~(2.1~\gev) and Belle
(2.25~\gev). Presently, experimental errors are only slightly larger
than the theoretical uncertainty. Computing a world average is
complicated by the correlated systematic and theoretical errors. The
dominant systematic error for the fully inclusive method is from the
\B\ decay background subtraction. The a dominant systematic error for
the semi--inclusive method is from the efficiencies of reconstructing
the final states, including a correction for final states not
considered in the analysis. This error is expected to be reduced
significantly in the next round of analyses. The average branching
fraction reported, ${\cal B} = (3.40 \pm 0.39)\times 10^{-4}$, is
computed assuming that the systematic errors are uncorrelated, for
simplicity. 

The present ${\cal B} ( B \rightarrow X_s \gamma)$ measurements
already provide a significant constraint on the SUSY parameter space.
For example limits on new physics contributions to $\B \rightarrow X_s
\gamma$ have been calculated using the minimal supergravity model
(SUGRA)~\cite{hewett} and charged Higgs bosons~\cite{misiak}.

So far, only CLEO~\cite{cleo3} has measured the direct \CP\ asymmetry.
Their technique does not suppress the background coming from $b\to d
\gamma$ decays (which is expected to have a large \CP\ asymmetry).
The measured direct \CP\ asymmetry is $0.965\times {\cal
A}_{CP}(\B \rightarrow X_s \gamma) + 0.02\times{\cal A}_{CP}(\B
\rightarrow X_d \gamma) = (-0.079 \pm 0.108 \pm 0.022) \times (1.0 \pm
0.03)$. The first error is statistical, while the second and third
errors represent additive and multiplicative systematic uncertainties,
respectively. The theoretical expectation of ${\cal B}(\B \rightarrow
X_d \gamma)$ is used. Results are consistent with no asymmetry.
 
\BaBar~\cite{BaBarbxs} and CLEO~\cite{CLEObsg} have published a
measurement of the photon energy spectrum down to a threshold
$E_{\gamma}^* > 2.1$ and 2.0~\gev, respectively, where $E_{\gamma}^*$
is measured in the \B and in the laboratory rest frame, respectively
(see Figure~\ref{fig:inclusive}). For the semi--inclusive analysis the
$E_\gamma$ spectrum is obtained from the mass of the hadronic system
$X_s$ because $E_{\gamma}$, because $E_{\gamma} = {{M^2_\B -
M^2_{X_s}}\over{2M_\B}}$ in the \B\ rest frame.

From the measured spectrum, \BaBar\ and CLEO have extracted the first
moment in the \B\ rest frame, $\langle E_\gamma \rangle$, finding
$\langle E_\gamma \rangle = 2.35 \pm 0.04 \pm 0.04$~\gev\ and $\langle
E_\gamma \rangle = 2.346 \pm 0.032 \pm 0.011$~\gev, respectively. Using
expressions in the $\overline {MS}$ renormalization scheme, to order
$1/M_B^3$ and order $\alpha_s^2 \beta_0$~\cite{ligeti}, \BaBar\ and
CLEO obtain $\bar \Lambda = 0.37 \pm 0.09 \pm 0.07 \pm 0.10$~\gev\ and
$\bar \Lambda = 0.35 \pm 0.08 \pm 0.10$~\gev\ from the first moment.
The errors are statistical, systematic (combined in the CLEO
measurement) and theoretical, respectively. Moreover, CLEO has used
their measured $\B \to X_s \gamma$ photon energy spectrum to determine
the light--cone shape function. Using this information, CLEO extracts
$| V_{ub}| = (4.08\pm 0.34 \pm 0.44 \pm 0.16\pm 0.24)\times
10^{-3}$~\cite{CLEO-endpoint}, where the first two uncertainties are
experimental and the last two are from theory.

\hspace{-5cm}
\hspace{-5cm}

%%%%%%%%%%%%%%%%%%%%%%%%%%%%%%%%%%%%%%%%%
\subsection{The Exclusive Process $\B\to K^*\gamma$}

For the exclusive decay, $B \rightarrow K^* \gamma$, two recent NLO
calculations predict SM branching fractions of ${\cal B} (B \rightarrow
K^* \gamma) = (7.1^{+2.5}_{-2.3}) \times 10^{-5}$ \cite{bosch} and
${\cal B} (B \rightarrow K^* \gamma) = (7.9^{+3.5}_{-3.0}) \times
10^{-5}$ \cite{beneke}. The errors are still dominated by the
uncertainties in the form factors.

The exclusive $B \rightarrow K^* \gamma$ modes have been studied by
\BaBar~\cite{babar1}, Belle~\cite{belle1} and CLEO~\cite{cleo1}, where
Belle used the highest statistics sample. Utilizing kinematic
constraints resulting from a full reconstruction of the $B$ decay,
substantial reduction of the $q \bar q$ (continuum) background is seen.
The Belle \emph{beam--constrained}
~\footnote{Results for exclusive \B\ decays are typically presented
using the following kinematic variables. If $(E_B^*, \vec{p}_B^*) $ is
the four--momentum of a reconstructed \B candidate in the overall CM
(\Y4S) frame, we define
\begin{center}
$ \de \equiv E_B^* - \ebeam \ ,$
 \label{eq:de}
\end{center}
\begin{center}
 $\mes\ (\mathrm{or}\ M_{bc}) \equiv \sqrt{{\ebeam}^2 - p_B^{*2}}\ .$
 \label{eq:mes}
\end{center}
The latter is called the \emph{energy--substituted} (\BaBar) or
\emph{beam--constrained} (Belle, CLEO) mass. Signal events peak at \de\
near 0~\gev\ and \mes ($M_{bc}$) near \B\ meson mass; whereas continuum
background lacks peaks.} 
mass distribution is shown in Figure~3 for all the $K^*$
decay channels. The measured branching fractions from all the
experiments and the corresponding average are summarized in
Table~\ref{tab:kstar}. The average branching fraction measurement is
consistent with the NLO SM predictions and is known to higher accuracy
than the theoretical uncertainty of $35$--$40\%$.

\begin{figure}[hbtp]
\begin{center}
\includegraphics[width=10cm,height=7cm]{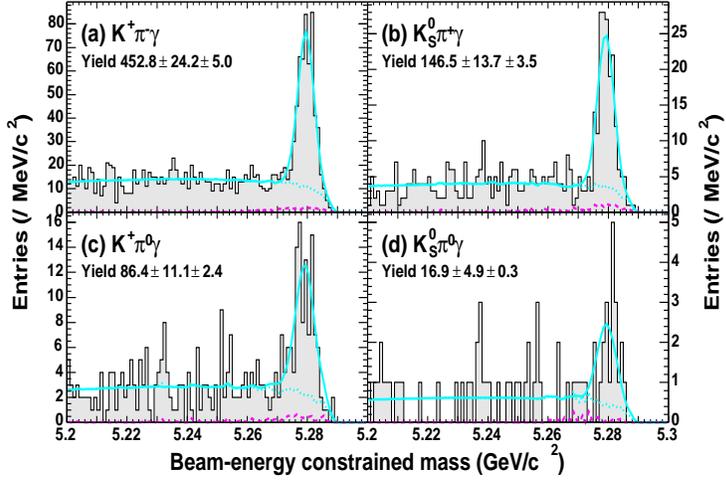}
\caption{Belle~\cite{belle1} \emph{beam--constrained} mass distribution for the 
exclusive $B\to K^*\gamma$ for the four $K^*$ final states.}
\end{center}
\label{fig:kstar}
\end{figure}

The direct \CP\ asymmetry is defined, at the quark level, as:
\begin{displaymath}
A_{CP} = {{\Gamma(b\to s \gamma ) - \Gamma(\overline{b}\to \overline{s} \gamma)}\over{\Gamma(b\to s \gamma ) + \Gamma(\overline{b}\to \overline{s} \gamma)}}.
\end{displaymath}
Table~\ref{tab:kstar} also summarizes the measurements of the direct \CP\ asymmetry in 
$B\to K^*\gamma$. These are consistent with zero and are statistics limited.

Isospin asymmetry, $\Delta_{0+}$, is calculated by Belle using the world average value 
$\tau_{\B^+}/\tau_{\B^0}$ = 1.083 $\pm$ 0.017~\cite{PDG}. The result is:
$\Delta_{0+} = + 0.003 \pm 0.045 \pm 0.018$, where the first error is statistical and the
second systematic. It is consistent with no asymmetry, having assumed equal production
of charged and neutral \B's at the $\Upsilon(4S)$ resonance. The isospin asymmetry 
can be used to set limits on Wilson coefficients~\cite{isospin}. 
\hspace{-1cm}
\begin{table}[htbp]
\begin{center}
\small
\begin{tabular}{|l|c|c|c|} \hline
                                                             & $B^0\to K^{*0}\gamma \times10^{-5}$ & $B^+\to K^{*+}\gamma \times10^{-5}$ & $A_{CP}$  \\ \hline
\BaBar~\cite{babar1} (21~\invfb)    &$4.23\pm 0.40\pm 0.22$                         &$3.83\pm 0.62\pm 0.22$                    & $-0.04\pm0.08\pm 0.01$\\
Belle~\cite{belle1} (78~\invfb)        &$4.09\pm 0.21\pm 0.19$&$4.40\pm 0.33\pm 0.24$                    & $-0.00\pm 0.04\pm 0.01$\\
CLEO~\cite{cleo1} (9~\invfb)          &$4.55{\,}^{+0.72}_{-0.68}\pm 0.34$&$3.76{\,}^{+0.89}_{-0.83}\pm 0.28$  & $+0.08\pm 0.13\pm 0.03$ \\ \hline
Average                                              &$4.18\pm 0.23$ & $4.14\pm 0.33$                 & $-0.01\pm 0.04$ \\ \hline
\end{tabular}
\end{center}
\caption{$B\to K^*\gamma$ branching fraction and direct \CP\ asymmetry measurements.}
\label{tab:kstar}
\end{table}

In addition to the already established 
$\B \to K^*\gamma$ decay, there are several known resonances that can contribute to the 
$X_s$ final state. Current measurements of higher than $K^*(892)$ mass systems are
from Belle~\cite{belle1430} and CLEO~\cite{cleo1}.
Note that the decay $\B\to \phi K \gamma$ was observed
recently by Belle~\cite{phiKgamma} for the first time. CLEO also 
observed radiative \B\ decays with baryons~\cite{CLEObaryons}.
Theoretical predictions cover a wide range; results so far are consistent with those 
from a relativistic form factor model as in Ref.~\cite{Veseli}.

%%%%%%%%%%%%%%%%%%%%%%%%%%%%%%%%%%%%%%%%%%%%%%%%%%%%%%%%%%%%%%%%%%%%%%%%%%%%%%%%%%%%%%%%%%%%%
\section{$b \to \d \gamma$ final states}

Both inclusive and exclusive $b \rightarrow d \gamma$ decays, which are 
suppressed by $\mid V_{td} / V_{ts} \mid^2 \sim 1/20$ with respect to corresponding
$b \rightarrow s \gamma$ modes, have not been seen yet.
An NLO calculation, which includes long--distance effects of $u$--quarks in the penguin loop,
predicts a range of
$ 6.0 \times 10^{-6} \leq {\cal B}(B \rightarrow X_d \gamma) \leq 2.6 \times
10^{-5}$~\cite{ali2} for the inclusive branching fraction. 
The uncertainty is dominated by imprecisely known CKM parameters. 

A measurement branching fraction ratio of ${\cal B}(B \rightarrow X_d
\gamma) / {\cal B}(B \rightarrow X_s \gamma)$ provides a determination
of $\mid V_{td} / V_{ts} \mid$ with small theoretical uncertainties. A
determination of $\mid V_{td} / V_{ts} \mid$ in the exclusive modes $B
\rightarrow \rho(\omega) \gamma$ has somewhat enhanced model
uncertainties, since form factors are not precisely known. The \CP\
asymmetry predicted in the SM for the inclusive process is foreseen
between $\sim 7~\%$ and $\sim 35~\%$~\cite{ali2}.

Studies of the $b\to d \gamma $ decays presently focus 
on searching for the exclusive process $\B\to \rho/\omega \gamma$.
The corresponding branching fraction is predicted to be
${\cal B} (B  \rightarrow \rho \gamma) = (1.6^{+0.8}_{-0.5}) \times 10^{-5}$
\cite{bosch}, while the \CP\ asymmetry is of the order of $10\%$ \cite{bosch}.

From the experimental point of view, the $\B\to \rho(\omega) \gamma$ 
is more difficult than $\B\to K^*\gamma$ because the backgrounds are bigger since this mode is
CKM suppressed and $u \bar u, d \bar d$ continuum processes are enhanced
compared to $s \bar s$ continuum processes.

The smallest upper limits on the exclusive decays $B\to \rho(\omega)\gamma$ 
come from \BaBar~\cite{BABARrhogamma}, which uses a neural network to 
suppress most of the continuum background. The $B\to K^*\gamma$ events 
are removed using particle identification to veto kaons, with a $K\to\pi$ 
fake rate of $\approx$1\%. A multi--dimensional likelihood fit is made 
to the remaining events to give 90~\% C.L. upper limits 
of 1.2, 2.1 and 1.0$\times 10^{-6}$ on $\rho^0\gamma$, $\rho^+\gamma$ 
and $\omega\gamma$, respectively. Assuming isospin symmetry, this gives a 
combined limit ${\cal B}(B\to\rho\gamma)<1.9\times 10^{-6}$ (90~\% C.L.).
Limits from Belle and CLEO can be found in Refs.~\cite{BELLErhogamma} and \cite{cleo1}, respectively.

Of particular theoretical interest is the ratio ${\cal
B}(\B\to\rho\gamma)$ to ${\cal B}(\B\to K^*\gamma)$ as most of the
theoretical uncertainty cancels and so it can be used to determine the
ratio $\mid V_{td} / V_{ts} \mid$.~\cite{Alirhogamma} The present
limit is $\mid V_{td} / V_{ts} \mid < 0.36$ at 90\% confidence level,
and is not as tight as the constraint from $B_s/B_d$ mixing. However,
New Physics may appear in different ways in penguin and mixing
diagrams, so it is important to measure it in both processes.

\section{$b \to s l^+ l^-$}

In the SM, three amplitudes contribute at leading order to the $b \to
s l^+ l^-$ decay: an electromagnetic penguin, a Z penguin, and a $W^+
W^-$ box diagram.  The presence of three SM electroweak amplitudes
makes $b \to s l^+ l^-$ more complex than $b \to s \gamma$, which
proceeds solely through the EM penguin.  The branching fraction for $B
\to K l^+ l^-$ is predicted to be $0.5 \times 10^{-6}$, and three
times that for $B \to K^* l^+ l^-$.\cite{ali2002}

Although the $B \to K l^+ l^-$ has lower branching fraction than $B
\to K^* \gamma$, there are several experimental advantages that can be
exploited. Tracking of leptons is more accurate than measuring photons
in the calorimeter. A control sample is provided by $B \to J/\Psi (\to
l^+ l^-) K$ events which are used by both experiments to understand
efficiencies and systematics.

\begin{figure}[hbtp]
\begin{center}
\includegraphics[height=5cm]{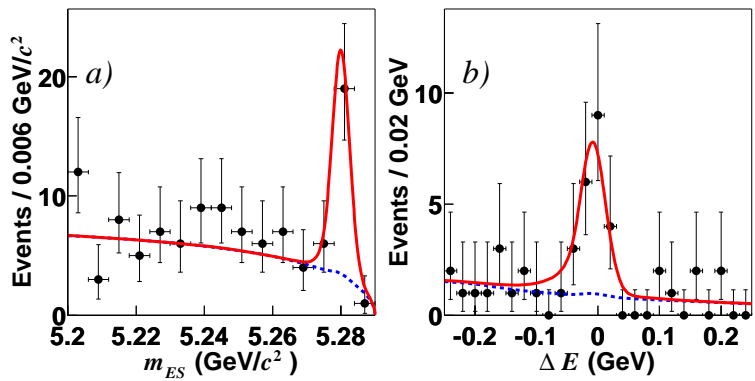}
\caption{\BaBar~\cite{BabarKll} energy-substituted mass and 
$\Delta E$ distributions (points) and projections (curves) of the simultaneous fit
used to extract the branching fraction results for the 
exclusive $B\to K l^+ l^-$ process are shown. (a) $m_{ES}$ for
$ -0.11 < \Delta E < 0.05$ GeV (b) $\Delta E$ for $|m_{ES} - m_B|<6.6$
MeV/c$^2$. The solid curve is the sum of signal and all background
fit components, whereas the dashed curve is just the sum of all
background fit components.}
\end{center}
\label{fig:kll}
\end{figure}

The Belle collaboration has observed $B \to K l^+ l^-$, as well as the
inclusive $B \to X_s l^+ l^-$ decay\cite{BelleKll, BelleXsll}, which follows the
``semi-inclusive'' $b \to s \gamma$ analysis strategies. The \BaBar\ 
collaboration has also observed $B \to K l^+ l^-$ and
$B \to X_s l^+ l^-$ decays\cite{BabarKll, BabarXsll}.
Figure~\ref{fig:kll} shows the \BaBar\  distributions of $M_{ES}$ and
$\Delta E$ along with the projections of the simultaneous fit used to
extract the branching fractions. For $K^* l^+ l^-$, the reconstructed
mass of $K^*$ is also used in the multidimensional fit. 

The electron and muon modes are
separately analysed. The lepton result is defined using 
$ {\cal B}(B\to K l^+ l^-) = {\cal B}( K \mu^+ \mu^-) = 0.75 \times {\cal B}(B \to K e^+ e^-)$, 
which accounts for larger electron $q^2 = 0$ pole. 
The measurements of the branching fractions are shown in
Table~\ref{tab:kll}. With more statistics, theoretically interesting
$M_{ll}$ spectrum can be measured.

\begin{table}[hbtp]
\begin{center}
\begin{tabular}{|l|c|c|c|} \hline
                                                             & $B^0\to K l^+ l^-$ & $B^+\to K^* l^+ l^- $ & $B\to X_s l^+ l^- $  \\ \hline
Belle~\cite{BelleKll,BelleXsll}       (130~\invfb)  & $0.48^{+0.10}_{-0.09} \pm 0.03$  &  $0.12^{+0.26}_{-0.24} \pm 0.08$  & $ 6.1 \pm 1.4^{+1.4}_{-1.1}$\\
\BaBar~\cite{BabarKll,BabarXsll} (113~\invfb)  & $0.65^{+0.14}_{-0.13} \pm 0.04$  &  $0.88^{+0.33}_{-0.29} \pm 0.10$  & $6.3 \pm 1.6^{+1.8}_{-1.5}$\\  \hline
\end{tabular}
\caption{Branching fractions in units of ($\times 10^{-6}$) for $B \to K l^+ l^-$, $B \to K^* l^+ l^-$  and
inclusive $B \to X_s l^+ l^-$.}
\label{tab:kll}
\end{center}
\end{table}

%%%%%%%%%%%%%%%%%%%%%%%%%%%%%%%%%%%%%%%%%%%%%%%%%%%%%%%%%%%%%%%%%%%%%%%%%%%%%%%%%%%%%%%%%%%
\section{Conclusions and Outlook}
A review of recent experimental results of radiative penguin decays
$b\to s(d)\gamma $ and $b \to s l^+ l^-$ is presented. The $b\to
s\gamma$ process both in the inclusive and exclusive final states is
well established. More statistics can be used to improve the limits or
indirectly find evidence of new physics in \CP\ and isospin
asymmetries. Higher statistics can also improve the measurement of the
photon energy spectrum, which is necessary to make better determination
of the CKM matrix elements. 

There is not yet any evidence of $b\to d\gamma$ decays but  
\BaBar\  and Belle expect to collect $\approx 500~\invfb$ by 2006.
This should be sufficient to observe $B\to\rho\gamma$ enabling
an important alternate method for the determination of $|V_{td}/V_{ts}|$. 
It may also be feasible to measure the inclusive $b\to d\gamma$ rate. For the measurement of 
$|V_{td}/V_{ts}|$, the ratio of $b\to d\gamma$ 
to $b\to s\gamma$ has much smaller theoretical uncertainties 
than the ratio of the exclusive decays.

The signals for $B \to K l^+ l^-$ are observed at 8$\sigma$ level, and
evidence for $B \to K^* l^+ l^-$ is seen by both Belle and \BaBar. So
far these results are in agreement with the Standard Model
expectations. 

%%%%%%%%%%%%%%%%%%%%%%%%%%%%%%%%%%%%%%%%%%%%%%%%%%%%%%%%%%%%%%%%%%%%%%%%%%%%%%%%%%%%%%%%%%%%

\section{Acknowledgements}
I would like to thank \BaBar, Belle and CLEO collaborations for making
the measurements available. I would also like to acknowledge the help
of Dr. F. Di Lodovico, University of Wisconsin, in preparing this
paper. This work is supported in part by grants from the U.S.
Department of Energy and the University of Wisconsin, WARF.

\end{document}

%% file: symbols.tex
\def\beq{\begin{equation}}
\def\eeq#1{\label{#1}\end{equation}}
\def\eeqn{\end{equation}}

\def\beqa{\begin{eqnarray}}
\def\eeqa#1{\label{#1}\end{eqnarray}}
\def\eeqan{\end{eqnarray}}

\let\bar=\overbar

\def\Dslash{\not{\hbox{\kern-4pt $D$}}}
\def\dslash{\not{\hbox{\kern-2pt $\del$}}}

\def\msb{{\bar{\ssstyle M \kern -1pt S}}}

\def\BB0bar{B^0 {\overline B}^0}
\def\BB0dbar{B_d^0 {\overline B}_d^0}
\def\BB0sbar{B_s^0 {\overline B}_s^0}

\RequirePackage{xspace}

\usepackage{relsize}
\def\babar{\mbox{\slshape B\kern-0.1em{\smaller A}\kern-0.1em
    B\kern-0.1em{\smaller A\kern-0.2em R}}}

\def\en         {\ensuremath{e^-}\xspace}   % electron negative (\em is taken)
\def\ep         {\ensuremath{e^+}\xspace}

\def\mup        {\ensuremath{\mu^+}\xspace}
\def\mun        {\ensuremath{\mu^-}\xspace} % muon negative (\mum is taken)
\def\mumu       {\ensuremath{\mu^+\mu^-}\xspace}

\def\taup       {\ensuremath{\tau^+}\xspace}
\def\taum       {\ensuremath{\tau^-}\xspace}

\def\ellell     {\ensuremath{\ell^+ \ell^-}\xspace}

\def\nub        {\ensuremath{\overline{\nu}}\xspace}
\def\nunub      {\ensuremath{\nu{\overline{\nu}}}\xspace}
\def\nub        {\ensuremath{\overline{\nu}}\xspace}
\def\nunub      {\ensuremath{\nu{\overline{\nu}}}\xspace}
\def\nue        {\ensuremath{\nu_e}\xspace}
\def\nueb       {\ensuremath{\nub_e}\xspace}

\def\num        {\ensuremath{\nu_\mu}\xspace}
\def\numb       {\ensuremath{\nub_\mu}\xspace}

\def\nut        {\ensuremath{\nu_\tau}\xspace}
\def\nutb       {\ensuremath{\nub_\tau}\xspace}

\def\nul        {\ensuremath{\nu_\ell}\xspace}
\def\nulb       {\ensuremath{\nub_\ell}\xspace}

\def\g     {\ensuremath{\gamma}\xspace}
\def\gaga  {\ensuremath{\gamma\gamma}\xspace}  %% changed from \gg, which is >>
\def\ggstar{\ensuremath{\gamma\gamma^*}\xspace}

\def\d     {\ensuremath{d}\xspace}

\def\b     {\ensuremath{b}\xspace}

\def\piz   {\ensuremath{\pi^0}\xspace}

\def\ppz   {\ensuremath{\pi^0\pi^0}\xspace}
\def\pip   {\ensuremath{\pi^+}\xspace}
\def\pim   {\ensuremath{\pi^-}\xspace}
\def\pipi  {\ensuremath{\pi^+\pi^-}\xspace}

\def\pimp  {\ensuremath{\pi^\mp}\xspace}

\def\kaon  {\ensuremath{K}\xspace}
\def\Kbar  {\kern 0.2em\overline{\kern -0.2em K}{}\xspace}

\def\Kz    {\ensuremath{K^0}\xspace}
\def\Kzb   {\ensuremath{\Kbar^0}\xspace}
\def\KzKzb {\ensuremath{\Kz \kern -0.16em \Kzb}\xspace}
\def\Kp    {\ensuremath{K^+}\xspace}
\def\Km    {\ensuremath{K^-}\xspace}
\def\Kpm   {\ensuremath{K^\pm}\xspace}

\def\KpKm  {\ensuremath{\Kp \kern -0.16em \Km}\xspace}
\def\KS    {\ensuremath{K^0_{\scriptscriptstyle S}}\xspace} 
\def\KL    {\ensuremath{K^0_{\scriptscriptstyle L}}\xspace} 
\def\Kstarz  {\ensuremath{K^{*0}}\xspace}

\def\Kstar   {\ensuremath{K^*}\xspace}

\def\Kstarp  {\ensuremath{K^{*+}}\xspace}

\def\Dbar    {\kern 0.2em\overline{\kern -0.2em D}{}\xspace}

\def\Dz      {\ensuremath{D^0}\xspace}
\def\Dzb     {\ensuremath{\Dbar^0}\xspace}
\def\DzDzb   {\ensuremath{\Dz {\kern -0.16em \Dzb}}\xspace}
\def\Dp      {\ensuremath{D^+}\xspace}
\def\Dm      {\ensuremath{D^-}\xspace}

\def\DpDm    {\ensuremath{\Dp {\kern -0.16em \Dm}}\xspace}
\def\Dstar   {\ensuremath{D^*}\xspace}

\def\Dstarz  {\ensuremath{D^{*0}}\xspace}

\def\Dstarp  {\ensuremath{D^{*+}}\xspace}
\def\Dstarm  {\ensuremath{D^{*-}}\xspace}

\def\B       {\ensuremath{B}\xspace}
\def\Bbar    {\kern 0.18em\overline{\kern -0.18em B}{}\xspace}

\def\BB      {\ensuremath{B\Bbar}\xspace} 
\def\Bz      {\ensuremath{B^0}\xspace}
\def\Bzb     {\ensuremath{\Bbar^0}\xspace}
\def\BzBzb   {\ensuremath{\Bz {\kern -0.16em \Bzb}}\xspace}
\def\Bu      {\ensuremath{B^+}\xspace}
\def\Bub     {\ensuremath{B^-}\xspace}

\def\Bpm     {\ensuremath{B^\pm}\xspace}

\def\BpBm    {\ensuremath{\Bu {\kern -0.16em \Bub}}\xspace}

\def\jpsi     {\ensuremath{{J\mskip -3mu/\mskip -2mu\psi\mskip 2mu}}\xspace}

\mathchardef\Upsilon="7107
\def\Y#1S{\ensuremath{\Upsilon{(#1S)}}\xspace}% no space before {...}!

\def\FourS {\Y4S}

\mathchardef\Deltares="7101
\mathchardef\Xi="7104
\mathchardef\Lambda="7103
\mathchardef\Sigma="7106
\mathchardef\Omega="710A

\def\Deltabar{\kern 0.25em\overline{\kern -0.25em \Deltares}{}\xspace}
\def\Lbar{\kern 0.2em\overline{\kern -0.2em\Lambda\kern 0.05em}\kern-0.05em{}\xspace}
\def\Sigbar{\kern 0.2em\overline{\kern -0.2em \Sigma}{}\xspace}
\def\Xibar{\kern 0.2em\overline{\kern -0.2em \Xi}{}\xspace}
\def\Obar{\kern 0.2em\overline{\kern -0.2em \Omega}{}\xspace}
\def\Nbar{\kern 0.2em\overline{\kern -0.2em N}{}\xspace}
\def\Xb{\kern 0.2em\overline{\kern -0.2em X}{}\xspace}

\def\X {\ensuremath{X}\xspace}

\def\BR         {{\ensuremath{\cal B}\xspace}}

\def\bpsikl     {\ensuremath{\Bz \to \jpsi \KL}\xspace}

\def\Bzbtox     {\ensuremath{\Bzb \to \X}\xspace}

\def\de         {\ensuremath {\Delta E^{*}}\xspace}

\def\mes        {\mbox{$m_{\rm ES}$}\xspace}

\def\ebeam      {\ensuremath {E^{*}_\mathrm{beam}}\xspace}

\def\invfb   {\ensuremath{\mbox{\,fb}^{-1}}\xspace}

\def\mus  {\ensuremath{\rm \,\mus}\xspace}

\def\mus        {\ensuremath{\,\mu{\rm s}}\xspace}    %% microsecond
\def\degrees{\ensuremath{^{\circ}}\xspace}
\def\to                 {\ensuremath{\rightarrow}\xspace}

\def\pep2{PEP-II}
\def\BF{$B$ Factory}